\documentclass[twocolumn,showpacs,preprintnumbers,amsmath,amssymb]{revtex4}
\usepackage[dvips]{graphicx}
\usepackage{dcolumn}
\usepackage{amsmath}
\usepackage{longtable}

\def\ketm#1{  \left\vert  #1   \right\rangle   }
\def\bram#1{  \left\langle  #1   \right\vert   }
\def\etal{\textit{et al.}}

\begin{document}

\title{Entanglement dynamics of three-qubit states in noisy channels}
\author{Michael Siomau}
\email{siomau@physi.uni-heidelberg.de}
\affiliation{Max-Planck-Institut f\"{u}r Kernphysik, Postfach
             103980, 69029 Heidelberg, Germany}
\affiliation{Physikalisches Institut, Heidelberg Universit\"{a}t,
69120 Heidelberg, Germany}
\author{Stephan Fritzsche}
\affiliation{Department of Physical Sciences, P.O.~Box 3000,
             90014 University of Oulu, Finland}
\affiliation{Frankfurt Institute for Advanced Studies,
             60438 Frankfurt am Main, Germany}

\date{\today}

\begin{abstract}
We study entanglement dynamics of the three-qubit system which is
initially prepared in pure Greenberger-Horne-Zeilinger (GHZ) or W
state and transmitted through one of the Pauli channels $\sigma_z,
\, \sigma_x, \, \sigma_y$ or the depolarizing channel. With the help
of the lower bound for three-qubit concurrence we show that the W
state preserves more entanglement than the GHZ state in transmission
through the Pauli channel $\sigma_z$. For the Pauli channels
$\sigma_x, \, \sigma_y$ and the depolarizing channel, however, the
entanglement of the GHZ state is more resistant against decoherence
than the W-type entanglement. We also briefly discuss how the
accuracy of the lower bound approximation depends on the rank of the
density matrix under consideration.
\end{abstract}


\maketitle

\section{\label{sec:1} Introduction}

Entanglement is the fundamental property of quantum systems which
has been found as a resource for many potential applications, such
as quantum teleportation, cryptography and superdense coding
\cite{Nielsen:00}, to name just a few. Besides the other
manifestations of entanglement, quantum teleportation still holds
central place in quantum information theory, due to a few remarkable
recent experimental achievements \cite{Roos:04, Fedrizzi:09, Lu:07}.
A protocol for quantum teleportation was originally suggested in the
seminal works of Ekert \cite{Ekert:91} and of Bennett \etal{}
\cite{Bennett:93} to facilitate communication between two partners.
The protocol enables one to transmit an unknown quantum state to the
remote recipient by means of an initially shared two-qubit entangled
(Bell's) state. Soon thereafter, communication protocols for quantum
teleportation between several partners, that are based on many-qubit
entangled states, were suggested \cite{Karlsson:98, Pati:06}. In
particular, Karlsson and Bourennane \cite{Karlsson:98} suggest
protocols for quantum teleportation between two and three partners
with three-qubit GHZ state
\begin{equation}
\label{pure-GHZ}
 \ketm{GHZ} = \frac{1}{\sqrt{2}} \left( \ketm{000} +
 \ketm{111} \right) \, .
\end{equation}
Later Agrawal and Pati suggested using three-qubit W state
\begin{equation}
\label{pure-W}
 \ketm{W} = \frac{1}{2} \left( \sqrt{2} \ketm{001} +
 \ketm{010} + \ketm{100} \right) \, ,
\end{equation}
in protocols for quantum teleportation between two partners and
superdense coding \cite{Pati:06}. Nowadays, protocols for quantum
teleportation, that are based on many-qubit entangled states, are
widely used in order to construct quantum networks
\cite{Neergaard:06, Kimble:08, Prevedel:09}.

While protocols for quantum teleportation require pure entangled
states, in practice one has to deal with entangled states that are
mixed due to imperfect control during their transmission through
communication channels. It is known that a pure entangled state can
be extracted from several copies of mixed entangled states, if a
so-called protocol for entanglement purification is applied to the
mixed states \cite{Bennett:96, Smolin:96, Vedral:98}. To provide the
protocol on received mixed states it is necessary to know how much
entanglement is preserved in a single mixed state after transmission
through a (noisy) communication channel.

The entanglement dynamics of many-qubit entangled states under
influence of different types of environment has been intensively
discussed during last decade \cite{Carvahlo:04, Mintert:05,
Konrad:07, Liu:09}. For instance, entanglement dynamics of
three-qubit GHZ and W states for special types of environment
coupling, such as a thermal bath at zero and infinite temperatures
as well as the dephasing (i.e. the Pauli $\sigma_z$) channel, was
discussed in Ref.~\cite{Carvahlo:04}. In particular, with the help
of the lower bound for three-qubit concurrence it was shown that
GHZ-type entanglement is more fragile under these types of
environment coupling than the entanglement of the W state.

In this contribution we also study the entanglement dynamics of
three-qubit GHZ (\ref{pure-GHZ}) and W (\ref{pure-W}) states under
the influence of its environment. We compliment results which were
obtained in Ref.~\cite{Carvahlo:04} by considering entanglement
evolution of the states (\ref{pure-GHZ})-(\ref{pure-W}) in
transmission through various noisy channels, namely, the Pauli
channels $\sigma_z, \sigma_x$ and $\sigma_y$ as well as the
depolarizing channel \cite{Nielsen:00}. Although the entanglement
dynamics of the three-qubit states in the Pauli channel $\sigma_z$
was considered earlier in Ref.~\cite{Carvahlo:04} we include this
channel in our discussion in order to complete the investigation of
the Pauli channels.

The time evolution of three-qubit (open) quantum systems which are
initially prepared in the pure states
(\ref{pure-GHZ})-(\ref{pure-W}) and transmitted through one of the
Pauli channels or the depolarizing channel was recently analyzed by
Jung \etal{} \cite{Jung:08}. For each of these channels, the time
evolution of the GHZ and W states is given by mixed state density
matrices which were obtained by analytical solving of corresponding
master equations in Lindblad form. With the help of the analytic
expressions for the mixed state density matrices we describe the
entanglement evolution of the mixed GHZ and W states. To quantify
the entanglement of the mixed states we use the lower bound for the
three-qubit concurrence which was recently suggested by Li \etal
\cite{Li:09}. We show that the W state preserves more entanglement
than the GHZ state in transmission through the Pauli channel
$\sigma_z$. Surprisingly, for the Pauli channels $\sigma_x$ and
$\sigma_y$ as well as the depolarizing channel GHZ-type entanglement
is more resistant against decoherence than W-type entanglement.

In addition, we briefly discuss how accuracy of the lower bound
approximation for three-qubit concurrence depends on the rank of the
density matrix. In general, to quantify entanglement of a
three-qubit mixed state, the convex roof extension for three-qubit
concurrence is already known \cite{Horodecki:09}. Unfortunately the
calculation of the convex roof implies an optimization procedure
which still has no analytic solution. Of course, the optimization
can be done numerically. However, the calculation of the convex roof
for a density matrix with rank $r$ leads to optimization over $r^3$
free parameters \cite{Mintert:05} which is indeed quite a formidable
task. In practice moreover, one is mostly interested in the minimal
(nontrivial) amount of entanglement which is preserved in a mixed
state. With regard to this practical requirement several analytical
lower bound approximations for the concurrence were recently
suggested \cite{Mintert:05, Li:09, Chen:05} and were justified to be
proper entanglement measures. It is important to know how accurate
the approximation is and from which parameters of the density matrix
this accuracy depends on? As we have already mentioned in this work
we use the lower bound approximation for the concurrence from
Ref.~\cite{Li:09} in order to describe entanglement evolution of the
mixed states (\ref{pure-GHZ})-(\ref{pure-W}). We found that for the
density matrices with rank $r \leq 4$ the lower bound coincides with
the convex roof for the concurrence. For the density matrices with
higher rank the application of the lower bound becomes restricted:
the lower bound can not describe long-time evolution of the mixed
state.

The paper is organized as follows. In the next section we present
the necessary theoretical background to describe the time evolution
of the states (\ref{pure-GHZ})-(\ref{pure-W}) in the Pauli channels
$\sigma_z, \sigma_x$ and $\sigma_y$ as well as the depolarizing
channel. In Subsection~\ref{subsec:2.2} we discuss the convex roof
extension for many-qubit concurrence and introduce the lower bound
to be used. With the help of presented technique we show the
entanglement dynamics of the GHZ and the W states in transmission
through the Pauli and the depolarizing channels in
Sections~\ref{sec:3} and \ref{sec:4} respectively. In
Section~\ref{sec:5} we summarize the results which were obtained in
the previous two sections. We also briefly discuss the relation
between the accuracy of the lower bound approximation and the rank
of the density matrix.

\section{\label{sec:2}Theory}

\subsection{\label{subsec:2.1}Time evolution of states transmitted through
                               noisy channels}

Decoherence is a well-known quantum phenomenon that occurs for all
open systems, i.e. if they are coupled to their environment. In
order to classify and better understand this (often undesired)
interaction of a system with its surroundings, a large number of
noise models has been investigated during the last decades
\cite{Breuer:02, Gardiner:00}, including various thermal bathes at
either zero or infinite temperature, the phase and the amplitude
damping, or just the Pauli channels $\sigma_\alpha \;\,(\alpha =
x,y,z)$, etc. Following similar lines, Jung \etal{} \cite{Jung:08}
have recently analyzed the time evolution of the three-qubit GHZ
(\ref{pure-GHZ}) and W (\ref{pure-W}) states, if they are
transmitted through one of the Pauli or the depolarizing channel, in
order to explore the efficiency of the `two-sided' teleportation
protocols that are based on these entangled states. In this work
\cite{Jung:08}, in more detail, an initially pure entangled state
$\rho (0)$ was supposed to be transmitted through (one of) these
channels for the time $t$, and its time evolution $\rho (t)$
obtained as solution of a (Lindblad-type) master equation
\begin{eqnarray}
\label{master-eqn} \frac{\partial \rho}{\partial t} = -i\left[ H_S,
\rho\right] + \sum_{i,\alpha} \left( L_{i, \alpha} \rho L^\dag_{i,
\alpha} - \frac{1}{2} \{L^\dag_{i, \alpha}L_{i, \alpha}, \rho\}
\right)  .
\end{eqnarray}
In this master equation, the (Lindblad) operators $L_{i, \alpha}$
were assumed to act independently upon the $i-$th qubit; for
example, the operator $L_{1, z} \equiv \sqrt{k}\sigma_z \otimes 1
\otimes 1$ describes the decoherence of the first qubit under a
phase-flip $\sigma_z$, and where the coupling constant $k$ is
approximately inverse to the \textit{decoherence time} with regard
to such a phase-flip. Later we shall also refer the Pauli channels
$\sigma_x$ and $\sigma_y$ to bit-flip and bit-phase-flip coupling of
the three-qubit system to the environment \cite{Nielsen:00}. For any
given Pauli channel $\sigma_\alpha$, therefore, the master equation
(\ref{master-eqn}) only includes three Lindblad operators, $L_{1,
\alpha},\, L_{2, \alpha}$ and $L_{3, \alpha}$, while nine of these
operators are needed for the depolarizing channel, $L_{i, \alpha},\;
(i=1,2,3,\; \alpha = x,y,z)$. In the latter case, each of the qubits
can be affected with equal coupling strength by all three Pauli
channels simultaneously \cite{Nielsen:00}.

\subsection{\label{subsec:2.2} Concurrence for three-qubit mixed states}

Knowing the time evolution of the three-qubit GHZ (\ref{pure-GHZ})
and W (\ref{pure-W}) states in transmission throw the Pauli and the
depolarizing channels, we still need to quantify the remaining
entanglement of the mixed states after they have been passed through
some of these noisy channels discussed above. For mixed states, in
fact, any exact quantification of their entanglement has been found
difficult, and no general solution is known until now
\cite{Horodecki:09} apart from two-qubit systems. In the latter
case, Wootters \textit{concurrence} \cite{Wootters:98} provides a
very powerful measure but, despite of its rather simple form for
just two qubits, there is not unique generalization of this measure
available even for mixed bipartite states, if the dimensions of the
associated Hilbert (sub-) spaces are larger than two
\cite{Coffman:00,Rungta:01,Chen:05}. The generalization of the
concurrence for mixed many-partite states still remains an open
problem. However, several useful lower bounds for the concurrence
have been proposed in the literature \cite{Coffman:00, Li:09,
Carvahlo:04, Mintert:05} in order to characterize the entanglement
of many-partite states. Here, we shall not discuss these suggested
`measures' in further detail but make use of a recent work by Li
\etal{} \cite{Li:09} who suggested the lower bound for the
concurrence for three-qubit states which is based on the positive
partial transpose and realignment separability criteria.

Before we shall further discuss the entanglement of mixed states,
let us first consider the concurrence of some \textit{pure} state
$\ketm{\psi}$. For any (pure) three-qubit state, for example, the
concurrence can be expressed as \cite{Li:09}
\begin{equation}
 \label{c3}
C_{3}(\ketm{\psi}) = \sqrt{\frac{1}{3} \left( 3 - {\rm Tr}\,
\rho_1^2 - {\rm Tr}\, \rho_2^2 - {\rm Tr}\, \rho_3^2 \right) } \, ,
\end{equation}
and where the reduced density matrices $\rho_i = {\rm
Tr}_{jk}\ketm{\psi}\bram{\psi}$ with $i \ne j \ne k$ are obtained by
tracing out the remaining two qubits. Using the definition
(\ref{c3}), we can easily calculate the concurrence $C_{3}(\ketm{\rm
GHZ}) = 1/\sqrt{2}$ and $C_{3}(\ketm{\rm W}) = \sqrt{3/8}$ for pure
GHZ (\ref{pure-GHZ}) and pure W (\ref{pure-W}) states which are
considered in the present work. Although both states are known to be
fully entangled \cite{Horodecki:09}, the definition (\ref{c3})
results in two different values; therefore, we shall re-normalize
the expression (\ref{c3}) for each state in such a way, that we have
$C_{3}(\ketm{\rm GHZ}) = C_{3}(\ketm{\rm W}) \,=\, 1$. This
re-normalization is justified from experimental viewpoint, since a
three-qubit maximally entangled state can be viewed as a single unit
of a quantum teleportation protocol.

Of course, any \textit{mixed} state can be expressed also as a
convex sum of some pure states $\{ \ketm{\psi_i} \}$: $\rho = \sum_i
\, p_i \ketm{\psi_i}\bram{\psi_i}$. Using this representation,
definition (\ref{c3}) for the entanglement of pure states can be
generalized for mixed states and gives rise to the so-called
\textit{convex roof (extension)} \cite{Horodecki:09}
\begin{equation}
 \label{def-of-Cmix}
C_3(\rho) = {\rm min} \sum_i p_i \, C_3(\ketm{\psi_i}) \, ,
\end{equation}
where the minimum has to be found with regard to \textit{all
possible} decompositions of $\rho$ into pure states
$\ketm{\psi_i}\:$ (and with positive coefficients $p_i$). The
maximal number $i_{\rm max}$ of pure states (in the ensemble) is
called the \textit{cardinality} of the decomposition. This
cardinality is not fixed by the rank $r$ of the density matrix,
though it is usually assumed to be $i_{\rm max} \le r^2$
\cite{Uhlmann:98, Mintert:05}. In practice, therefore, one would
need to search for all decompositions of $\rho$ into up to $r^2$
pure states in order to just evaluate the convex roof for a
three-qubit density matrix, implying an optimization procedure with
$r \times r^2 \,\sim\, r^3$ free parameters \cite{Mintert:05}. This
is indeed quite a formidable task already for three qubits, and the
situation becomes worse since no numerical algorithm could guarantee
to find the global minimum for the expression on the rhs of
Eq.~(\ref{def-of-Cmix}). An analytical solution is known for this
optimization task but refer to a very special case
\cite{Lohmayer:06}.

Instead of finding the exact minimum for the three-qubit concurrence
(\ref{def-of-Cmix}), the rather simple lower bound to this measure
was suggested by Li \etal{} \cite{Li:09}
\begin{equation}
\label{low-bound}
 \tau_3 (\rho) \equiv \sqrt{\frac{1}{3}\, \sum_{i=1}^6 \, (C_i^{12|3})^2
 + (C_i^{13|2})^2 + (C_i^{23|1})^2 } \, ,
\end{equation}
which is given in terms of three bipartite concurrences that
correspond to possible bipartite cuts of the three-qubit system. The
bipartite concurrence was originally introduced in Ref.~\cite{Ou:08}
and was proved to be an entanglement measure. The bipartite
concurrence $C^{12|3}$ for qubits $12$ and $3$ is given by sum of
six terms $C_i^{12|3}$ in Eqn.~(\ref{low-bound}); each term is
expressed as
\begin{equation}
 \label{concurence}
C_i^{12|3} = {\rm max} \{ 0, \lambda_i^1 - \lambda_i^2 - \lambda_i^3
- \lambda_i^4 \} \, ,
\end{equation}
where the $\lambda_i^k, \, k=1..4$ (for given $i$) are the square
roots of the four nonvanishing eigenvalues in decreasing order of
the matrix $\rho\: \tilde{\rho}_i^{12|3}$. These matrices are not
hermitian and are formed by the density matrix $\rho$ and its
complex conjugate $\rho^*$, and which is further transformed by the
operators $\{ S_i^{12|3} = L^{12}_i \otimes L^3_0,\; i=1...6 \}$ as:
$\tilde{\rho}_i^{12|3} = S_i^{12|3} \rho^\ast S_i^{12|3}$. Details
about the construction of these operators and the mathematics behind
can be found in Ref.~\cite{Li:09}. In this notation, moreover,
$L^3_0$ is the single generator of the group SO(2), while the
$L^{12}_i$ are the six generators of SO(4). When the generator
$L^3_0$ simply equal to the second Pauli matrix $\sigma_y$, the
generators $L^{12}_i$ can be expressed by means of the totally
antisymmetric Levi-Cevita symbol in four dimensions,
i.e.~$(L_{kl})_{mn} = - i \varepsilon_{klmn}; \; k,l,m,n =1..4$
\cite{Jones:98, Bohm:79}. The bipartite concurrences $C^{13|2}$ and
$C^{23|1}$ are defined in the same way as above.

With these remarks about the lower bound for the concurrence
(\ref{c3}), we now have two possibilities to describe the time
evolution of the entanglement $C_3 (\rho(t))$ if an initial state
$\rho(0)$ is passed through a noisy channel. To determine the convex
roof (\ref{def-of-Cmix}) as function of time $t$, we would first
need to split time $t$ into a set of steps, $0 < t_1 < ... < t_N$,
and optimize at each time step $C_3(\rho(t_i))$ with regard to $\sim
r^3$ parameters, where $r$ is the rank of the density matrix.
Finally, an interpolation for all times `in between' these steps
need to be performed. If, in contrast, we make use of the lower
bound (\ref{low-bound}), we can evaluate this bound for the density
matrix $\rho(t)$ analytically for all times $t$. In the next two
sections, we shall therefore apply the lower bound (\ref{low-bound})
to the concurrence $\tau_3(\rho(t))$ to analyze the decay of the
entanglement for an initially pure GHZ (\ref{pure-GHZ}) and pure W
state (\ref{pure-W}) separately. However, since the lower bound
(\ref{low-bound}) is only an approximation for the convex roof
(\ref{def-of-Cmix}), we shall explore the validity of this approach
if the density matrix $\rho(t)$ departs more and more from a pure
state due to its interaction with the noisy channels. In rather
simple case when the density matrix has rank $r \leq 4$ we shall
compare the lower bound for the concurrence with actual value of the
convex roof obtained numerically. In Section~\ref{sec:5}, we discuss
how the accuracy of this lower-bound approximation is related to the
rank of the density matrix which differs in dependence of the
initial state and considered channel.

\section{\label{sec:3}~Entanglement evolution of an initial GHZ
                       state under noise}

\subsection{\label{subsec:3.1} Pauli channel $\sigma_z$}

If an initially pure GHZ state (\ref{pure-GHZ}) is transmitted
through the Pauli channel $\sigma_z$, its time evolution is obtained
as solution of the master equation (\ref{master-eqn}) with Lindblad
operators $\left( L_{1,z}, L_{2,z}, L_{3,z} \right)$ and can be
expressed in terms of the rank-2 density matrix \cite{Jung:08}
\begin{eqnarray}
 \label{phase-flip-GHZ}
\rho (t) & = & \frac{1}{2} \left( \ketm{000}\bram{000} +
\ketm{111}\bram{111}    \right)
\\[0.1cm]\nonumber
  & + & \frac{1}{2} e^{-6 k t} \left( \ketm{000}\bram{111} +
\ketm{111}\bram{000} \right) \, .
\end{eqnarray}
For this mixed state, the lower bound (\ref{low-bound}) to the
three-qubit concurrence is a monoexponential function of time,
\begin{eqnarray}
 \label{pf-GHZ-conc}
\tau_3\left(\rho (t)\right) = e^{-6 k t} \, .
\end{eqnarray}
Since the rank of the density matrix (\ref{phase-flip-GHZ}) is two,
the convex roof extension (\ref{def-of-Cmix}) for this density
matrix can be even calculated analytically \cite{Carvahlo:04}. In
this case, the convex roof is shown to follow the behavior of the
nondiagonal elements (up to the normalization factor). In fact, the
convex roof for the density matrix (\ref{phase-flip-GHZ}) coincides
with the lower bound (\ref{pf-GHZ-conc}).

\subsection{ \label{subsec:3.2} Pauli channel $\sigma_x$}

If the GHZ state (\ref{pure-GHZ}) is instead transmitted through the
Pauli channel $\sigma_x$, its time evolution is given by the rank-4
density matrix \cite{Jung:08}
\begin{equation}
 \label{bit-flip-GHZ}
 \rho (t) = \frac{1}{8} \left(
 \begin{array}{cccccccc}
  \alpha_+ & 0 & 0 & 0 & 0 & 0 & 0 & \alpha_+ \\
  0 & \alpha_- & 0 & 0 & 0 & 0 & \alpha_- & 0 \\
  0 & 0 & \alpha_- & 0 & 0 & \alpha_- & 0 & 0 \\
  0 & 0 & 0 & \alpha_- & \alpha_- & 0 & 0 & 0 \\
  0 & 0 & 0 & \alpha_- & \alpha_- & 0 & 0 & 0 \\
  0 & 0 & \alpha_- & 0 & 0 & \alpha_- & 0 & 0 \\
  0 & \alpha_- & 0 & 0 & 0 & 0 & \alpha_- & 0 \\
  \alpha_+ & 0 & 0 & 0 & 0 & 0 & 0 & \alpha_+ \\
 \end{array} \right) \, ,
\end{equation}
with
\begin{eqnarray}
\label{def-bf-not} \nonumber
 \alpha_+  & = &  1 + 3 e^{-4 k t} \quad \mbox{and} \quad
 \alpha_- \; = \; 1 - e^{-4 k t} \, .
\end{eqnarray}
In this case, the lower bound (\ref{low-bound}) to the three-qubit
concurrence becomes
\begin{eqnarray}
\label{bf GHZ conc}
 \tau_3\left(\rho (t)\right) = e^{-4 k t} \, ,
\end{eqnarray}
i.e. the entanglement of the initial state decays less quickly for a
bit-flip coupling of the three qubits to the environment than for a
phase-flip.

For the rank-4 density matrix (\ref{bit-flip-GHZ}) we also
calculated numerically the convex roof (\ref{def-of-Cmix}). We found
that the lower bound (\ref{bf GHZ conc}) coincides with the
numerical values of the convex roof.

\subsection{ \label{subsec:3.3} Pauli channel $\sigma_y$}

For a transmission of the GHZ (\ref{pure-GHZ}) state through the
Pauli channel $\sigma_y$, the density matrix \cite{Jung:08}
\begin{equation}
\begin{small}
 \label{bit-phase-flip-GHZ}
 \rho(t) = \frac{1}{8} \left(
 \begin{array}{cccccccc}
  \alpha_+ & 0 & 0 & 0 & 0 & 0 & 0 & \beta_1 \\
  0 & \alpha_- & 0 & 0 & 0 & 0 & -\beta_2 & 0 \\
  0 & 0 & \alpha_- & 0 & 0 & -\beta_2 & 0 & 0 \\
  0 & 0 & 0 & \alpha_- & -\beta_2 & 0 & 0 & 0 \\
  0 & 0 & 0 & -\beta_2 & \alpha_- & 0 & 0 & 0 \\
  0 & 0 & -\beta_2 & 0 & 0 & \alpha_- & 0 & 0 \\
  0 & -\beta_2 & 0 & 0 & 0 & 0 & \alpha_- & 0 \\
  \beta_1 & 0 & 0 & 0 & 0 & 0 & 0 & \alpha_+ \\
 \end{array} \right) \, ,
\end{small}
\end{equation}
has full rank (i.e. rank 8), with the two functions
\begin{eqnarray}
\nonumber
 \beta_1 & = & 3 e^{-2 k t} + e^{-6 k t} \quad \mbox{and} \quad
 \beta_2 \;=\; e^{-2 k t} - e^{-6 k t} \, ,
\end{eqnarray}
respectively. For this matrix, the lower bound (\ref{low-bound}) to
the concurrence gives rise to
\begin{small}
\begin{eqnarray}
\label{bpf-GHZ-conc}
 \tau_3\left(\rho(t) \right) = {\rm max} \{ 0,
\: \frac{1}{4} \left( 3 e^{-2 k t} + e^{-4 k t} + e^{-6 k t} - 1
\right) \, \} \, ,
\end{eqnarray}
\end{small}
or, in other words, this lower bound vanishes already after some
finite time. Using the positive partial transpose separability
criteria \cite{Horodecki:09}, we verified however that the state
(\ref{bit-phase-flip-GHZ}) becomes separable only asymptotically for
$t \rightarrow \infty$, which implies that the lower bound
(\ref{bpf-GHZ-conc}) does not describe the long-term behavior of the
entanglement of an initial GHZ state if its is affected by
bit-phase-flip noise.

For the rank-8 density matrix (\ref{bit-phase-flip-GHZ}) the
numerical calculation of the convex roof (\ref{def-of-Cmix})
requires optimization over $8^3 = 512$ free parameters. The
numerical value of the convex roof (\ref{def-of-Cmix}) for the
rank-8 density matrix (\ref{bit-phase-flip-GHZ}) as well as for
other rank-8 density matrices discussed below has not been obtained
by us yet.

\subsection{ \label{subsec:3.4} Depolarizing channel}

If the state (\ref{pure-GHZ}) is transmitted through the
depolarizing channel, its density matrix has also rank-8 and takes
the form \cite{Jung:08}
\begin{equation}
 \label{iso GHZ}
 \rho(t) = \frac{1}{8} \left(
 \begin{array}{cccccccc}
  \tilde{\alpha}_+ & 0 & 0 & 0 & 0 & 0 & 0 & \gamma \\
  0 & \tilde{\alpha}_- & 0 & 0 & 0 & 0 & 0 & 0 \\
  0 & 0 & \tilde{\alpha}_- & 0 & 0 & 0 & 0 & 0 \\
  0 & 0 & 0 & \tilde{\alpha}_- & 0 & 0 & 0 & 0 \\
  0 & 0 & 0 & 0 & \tilde{\alpha}_- & 0 & 0 & 0 \\
  0 & 0 & 0 & 0 & 0 & \tilde{\alpha}_- & 0 & 0 \\
  0 & 0 & 0 & 0 & 0 & 0 & \tilde{\alpha}_- & 0 \\
  \gamma & 0 & 0 & 0 & 0 & 0 & 0 & \tilde{\alpha}_+ \\
 \end{array} \right) \, ,
\end{equation}
with
\begin{small}
\begin{eqnarray}
\nonumber
 \tilde{\alpha}_+ & = & 1 + 3 e^{-8 k t}, \; \;
 \tilde{\alpha}_- \;=\; 1 - e^{-8 k t} \; \; \mbox{and}
 \; \gamma  = 4 e^{-12 k t} \, .
\end{eqnarray}
\end{small}
Here, again, the lower bound (\ref{low-bound}) to the entanglement
vanishes already after some finite time due to the condition
\begin{eqnarray}
\label{dep GHZ conc}
 \tau_3\left(\rho(t)\right) = {\rm max} \{ 0, \:
 \frac{1}{4} \left( 4 e^{-12 k t} + e^{-8 k t} - 1 \right) \} \, .
\end{eqnarray}

\begin{figure}
\begin{center}
\includegraphics[scale=0.7]{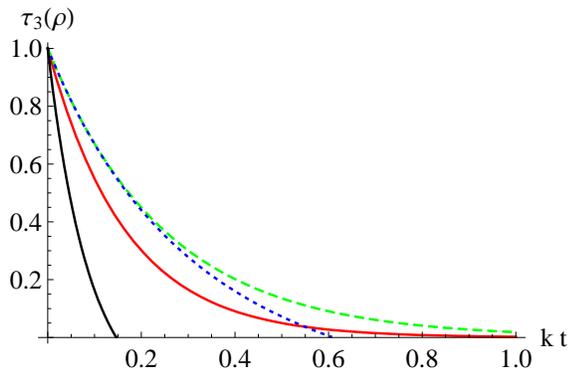}
\caption{(Color online) The lower bound (\ref{low-bound}) for the
three-qubit concurrence $\tau_3$ as function of time $t$ for an
initial GHZ state (\ref{pure-GHZ}), if transmitted through various
noisy channels: Pauli channels $\sigma_z$ (solid red), $\sigma_x$
(dashed green), $\sigma_y$ (dotted blue) and the depolarizing
channel (solid black).}
 \label{fig-1}
\end{center}
\end{figure}

Fig.~\ref{fig-1} displays the time-dependent lower bound
(\ref{low-bound}) for initial GHZ state (\ref{pure-GHZ}) if it's
transmitted through the different channels. In all cases, this lower
bound decays exponentially due to the noise of the channel; in
transmission throw the Pauli channels $\sigma_x$ and $\sigma_y$ the
entanglement of the GHZ state decreases slowly comparing to the
Pauli channels $\sigma_z$. The depolarizing coupling of the
three-qubit system to the channel is the most destructive for the
entanglement. It is also remarkable that for density matrices with
rank-2 and rank-4, the lower bound coincides with the convex roof
and describes the entanglement evolution for all times, while this
bound is not applicable for the long-time description of density
matrices with rank-8 (the Pauli $\sigma_y$ and the depolarizing
channels) for which it vanishes at a finite time.

\section{ \label{sec:4}~Entanglement evolution of an initial W
                       state under noise}

\subsection{ \label{subsec:4.1} Pauli channel $\sigma_z$}

A similar analysis as in Section~\ref{sec:3} can be made if the
system is initially prepared in a W state. If the state
(\ref{pure-W}) is transmitted through the channel $\sigma_z$, its
time evolution is described by the rank-three density matrix
\cite{Jung:08}
\begin{equation}
\begin{small}
 \label{phase flip W}
 \rho(t) = \frac{1}{4} \left(
 \begin{array}{cccccccc}
  0 & 0 & 0 & 0 & 0 & 0 & 0 & 0 \\
  0 & 2 & \sqrt{2} e^{-4 k t} & 0 & \sqrt{2} e^{-4 k t} & 0 & 0 & 0 \\
  0 & \sqrt{2} e^{-4 k t} & 1 & 0 & e^{-4 k t} & 0 & 0 & 0 \\
  0 & 0 & 0 & 0 & 0 & 0 & 0 & 0 \\
  0 & \sqrt{2} e^{-4 k t} & e^{-4 k t} & 0 & 1 & 0 & 0 & 0 \\
  0 & 0 & 0 & 0 & 0 & 0 & 0 & 0 \\
  0 & 0 & 0 & 0 & 0 & 0 & 0 & 0 \\
  0 & 0 & 0 & 0 & 0 & 0 & 0 & 0 \\
 \end{array} \right) \, ,
\end{small}
\end{equation}
and this gives rise to the lower bound
\begin{eqnarray}
 \label{pf-W-conc}
\tau_3\left(\rho(t)\right) = e^{-4 kt} \,
\end{eqnarray}
for the evolution of the entanglement, which moreover coincides with
the convex roof (\ref{def-of-Cmix}) as we verified numerically.

\subsection{ \label{subsec:4.2} Pauli channels $\sigma_x$ and $\sigma_y$}

If the (initially prepared) W state is transmitted through the Pauli
channels $\sigma_x$ or $\sigma_y$, a full rank-8 density matrix is
obtained for its time evolution \cite{Jung:08}
\begin{widetext}
\begin{small}
\begin{equation}
 \label{bit-phase-flip-W}
 \rho(t)_\pm = \frac{1}{16} \left(
 \begin{array}{cccccccc}
  2\alpha_2 & 0 & 0 & \pm\sqrt{2}\alpha_2 & 0
  & \pm\sqrt{2}\alpha_2 & \pm\alpha_2 & 0 \\
  0 & 2\alpha_1 & \sqrt{2}\alpha_1 & 0
  & \sqrt{2}\alpha_1 & 0 & 0 & \pm\alpha_3 \\
  0 & \sqrt{2}\alpha_1 & 2\beta_+ & 0 & \alpha_1
  & 0 & 0 & \pm\sqrt{2}\alpha_3 \\
  \pm\sqrt{2}\alpha_2 & 0 & 0 & 2\beta_- & 0
  & \alpha_4 & \sqrt{2}\alpha_4 & 0 \\
  0 & \sqrt{2}\alpha_1 & \alpha_1 & 0 & 2\beta_+ & 0
  & 0 & \pm\sqrt{2}\alpha_3 \\
  \pm\sqrt{2}\alpha_2 & 0 & 0 & \alpha_4 & 0
  & 2\beta_- & \sqrt{2}\alpha_4 & 0 \\
  \pm\alpha_2 & 0 & 0 & \sqrt{2}\alpha_4 & 0
  & \sqrt{2}\alpha_4 & 2\alpha_4 & 0 \\
  0 & \pm\alpha_3 & \pm\sqrt{2}\alpha_3 & 0
  & \pm\sqrt{2}\alpha_3 & 0 & 0 & 2\alpha_3 \\
 \end{array} \right) \, ,
\end{equation}
\end{small}
\end{widetext}
and where the $+$ sign refers to the $\sigma_x$ and $-$ to the
$\sigma_y$ channel, respectively. The time-dependent parameters in
expression (\ref{bit-phase-flip-W}) are given by
\begin{eqnarray}
\nonumber
 \alpha_1 & = & 1 + e^{-2 k t} + e^{-4 k t} + e^{-6 k t} \,
 \\[0.1cm] \nonumber
 \alpha_2 & = & 1 + e^{-2 k t} - e^{-4 k t} - e^{-6 k t} \,
 \\[0.1cm] \nonumber
 \alpha_3 & = & 1 - e^{-2 k t} - e^{-4 k t} + e^{-6 k t} \,
 \\[0.1cm] \nonumber
 \alpha_4 & = & 1 - e^{-2 k t} + e^{-4 k t} - e^{-6 k t} \; \;
 \mbox{and} \; \;
 \beta_\pm  =  1 \pm e^{-6 k t} \, .
\end{eqnarray}
Since two density matrices $\rho(t)_\pm$ have the same structure of
matrix elements, the lower bounds for these density matrices
coincide. Unfortunately, achieved analytic expression for the lower
bound for the density matrix (\ref{bit-phase-flip-W}) has no compact
form and, thus, we do not show it here explicitly. At
Fig.~\ref{fig-2} the lower bound is shown with blue dashed line. As
for all rank-8 density matrices above the lower bound for the
density matrix (\ref{bit-phase-flip-W}) vanishes after finite time.

\subsection{ \label{subsec:4.3} Depolarizing channel}

Finally, if the W state (\ref{pure-W}) is transmitted through the
depolarizing channel, the density matrix $\rho(t)$ has also rank-8
and is given by \cite{Jung:08}
\begin{small}
\begin{equation}
 \label{iso W}
 \frac{1}{8} \left(
 \begin{array}{cccccccc}
  \tilde{\alpha}_2 & 0 & 0 & 0 & 0 & 0 & 0 & 0 \\
  0 & \tilde{\alpha}_1 & \sqrt{2}\tilde{\gamma}_+
  & 0 & \sqrt{2}\tilde{\gamma}_+ & 0 & 0 & 0 \\
  0 & \sqrt{2}\tilde{\gamma}_+ & \tilde{\beta}_+
  & 0 & \tilde{\gamma}_+ & 0 & 0 & 0 \\
  0 & 0 & 0 & \tilde{\beta}_- & 0 & \tilde{\gamma}_-
  & \sqrt{2}\tilde{\gamma}_- & 0 \\
  0 & \sqrt{2}\tilde{\gamma}_+ & \tilde{\gamma}_+
  & 0 & \tilde{\beta}_+ & 0 & 0 & 0 \\
  0 & 0 & 0 & \tilde{\gamma}_- & 0 & \tilde{\beta}_-
  & \sqrt{2}\tilde{\gamma}_- & 0 \\
  0 & 0 & 0 & \sqrt{2}\tilde{\gamma}_- & 0
  & \sqrt{2}\tilde{\gamma}_- & \tilde{\alpha}_4 & 0 \\
  0 & 0 & 0 & 0 & 0 & 0 & 0 & \tilde{\alpha}_3 \\
 \end{array} \right) \, ,
\end{equation}
\end{small}
where
\begin{eqnarray}
\nonumber
 \tilde{\alpha}_1 &=& 1 + e^{-4 k t} + e^{-8 k t} + e^{-12 k t} \, ,
\\[0.1cm] \nonumber
 \tilde{\alpha}_2 &=& 1 + e^{-4 k t} - e^{-8 k t} - e^{-12 k t} \, ,
\\[0.1cm] \nonumber
 \tilde{\alpha}_3 &=& 1 - e^{-4 k t} - e^{-8 k t} + e^{-12 k t} \, ,
\\[0.1cm] \nonumber
 \tilde{\alpha}_4 &=& 1 - e^{-4 k t} + e^{-8 k t} - e^{-12 k t} \, ,
\\[0.1cm] \nonumber
 \tilde{\beta}_\pm &=& 1 \pm e^{-12 k t} \; \; \mbox{and} \; \;
 \tilde{\gamma}_\pm = e^{-8 k t} \pm e^{-12 k t} \, .
\end{eqnarray}

\begin{figure}
\begin{center}
\includegraphics[scale=0.7]{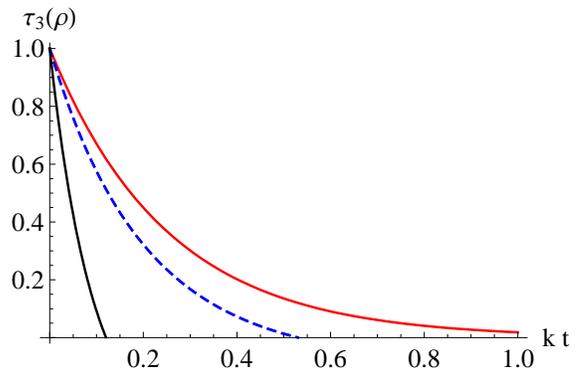}
\caption{(Color online) The same as in Fig.~\ref{fig-1} but for an
initial W state (\ref{pure-W}); the lower bound (\ref{low-bound})
for the three-qubit concurrence $\tau_3$ as function of time $t$ is
shown for noisy channels: Pauli channels $\sigma_z$ (solid red),
$\sigma_x$ and $\sigma_y$ (dashed blue) as well as the depolarizing
channel (solid black)}
 \label{fig-2}
\end{center}
\end{figure}

The time-dependent lower bound (\ref{low-bound}) for initial W state
(\ref{pure-W}) transmitted through the different channels is shown
at Fig.~\ref{fig-2}. As in the case of the GHZ state the lower
bounds for the W state decay exponentially due to the noise of the
channels. In contrast to the GHZ state, the entanglement of the W
state decreases slowly in transmission throw the Pauli channel
$\sigma_z$ comparing to the Pauli channels $\sigma_x$ and
$\sigma_y$. However, the depolarizing coupling of the three-qubit
system to the channel is again the most destructive for the
entanglement. For the rank-3 density matrix, moreover, the lower
bound coincides with the convex roof and describes the time
evolution of the entanglement for all times, while this bound is not
suitable for the long-time description of density matrices with rank
eight (the Pauli $\sigma_x$ and $\sigma_y$ as well as depolarizing
channels).

\section{ \label{sec:5} Results and discussion}

In previous two sections we showed explicitly the entanglement
evolution of the three-qubit system which is prepared in pure GHZ
(\ref{pure-GHZ}) or W (\ref{pure-W}) state and transmitted for the
time $t$ through one of the Pauli channels $\sigma_z$, \, $\sigma_x,
\, \sigma_y$ or the depolarizing channel. Having these results we
can investigate entanglement of which state between the GHZ and the
W is more resistant against decoherence in transmission through the
noisy channels. For the Pauli channel $\sigma_z$ the lower bounds
for the GHZ and the W states are given by Eqns.~(\ref{pf-GHZ-conc})
and (\ref{pf-W-conc}) respectively and are shown with red solid
lines at Figs.~\ref{fig-1}-\ref{fig-2}. In fact, the W state
preserves more entanglement than the GHZ state in transmission
through the Pauli channel $\sigma_z$ for all times $t$. This result
was obtained earlier in Ref.~\cite{Carvahlo:04}. For the Pauli
channels $\sigma_x$, \, $\sigma_y$ and the depolarizing channel, in
contrast, the lower bounds for the GHZ state, that are shown at
Fig.~\ref{fig-1}, are always higher than corresponding lower bounds
for the W state at Fig.~\ref{fig-2}. The entanglement of the GHZ
state is thus more resistant against decoherence than W-type
entanglement in transmission through the Pauli channels $\sigma_x$,
\, $\sigma_y$ and the depolarizing channel.

Our result extend the investigation of the entanglement dynamics of
the three-qubit GHZ and W states in transmission through noisy
channels, which was started in Ref.~\cite{Carvahlo:04}. As it has
been already known, the W state preserves more entanglement than the
GHZ state if it is coupled to the thermal bath at zero or infinite
temperature, or the Pauli channel $\sigma_z$ \cite{Carvahlo:04}. We
showed that the W state preserves less entanglement than the GHZ
state if it is coupled to the Pauli channels $\sigma_x$ or
$\sigma_y$, or the depolarizing channel.

To describe entanglement evolution of the mixed GHZ and W states we
used the lower bound (\ref{low-bound}) which is an approximation for
the convex roof for three-qubit concurrence (\ref{def-of-Cmix}). We
showed that the lower bound coincides with the convex roof for the
density matrices with rank $r \leq 4$. This statement, however, is
true only for the density matrices which were considered in
Subsections~\ref{subsec:3.1},\, \ref{subsec:3.2} and
\ref{subsec:4.1}. Whether the lower bound coincides with the convex
roof for an arbitrary density matrix with rank $r \leq 4$ introduces
an open question of great importance. We like to investigate this
question in the nearest future.

For the rank-8 density matrices discussed in this work the lower
bound for the concurrence vanishes after the finite time. Vanishing
of the lower bound, however, does not mean vanishing of the
entanglement of the mixed states. We verified with the positive
partial transpose separability criteria \cite{Horodecki:09} that the
mixed states become separable only asymptotically for the limit $t
\rightarrow \infty$. For the rank-8 density matrices the lower bound
for the concurrence can describe entanglement evolution of the mixed
state for a restricted time $t < t_0$ only, where $t_0$ denotes the
time when the lower bound vanishes. The comparison of the lower
bound with the convex roof for the rank-8 density matrices has not
been yet performed, since the calculation of the convex roof
introduce quite difficult problem which requires the numerical
optimization over $8^3 = 512$ free parameters. This comparison is
left for future research.

The lower bound approximation for the concurrence introduces a
powerful simple tool to describe the entanglement of an arbitrary
N-partite mixed state. Systematic investigation of the accuracy of
the approximation has not already been done. The fact that the
accuracy of the lower bound depends on the rank of the density
matrix is intuitively understandable, since the complexity of the
calculation of the convex roof also depends on the rank. In this
work we started investigation of the accuracy of the lower bound
approximation with regard to the rank of the density matrix. On
particular examples we showed that the lower bound
(\ref{low-bound}): \textsl{i.} coincides with the convex roof
(\ref{def-of-Cmix}) for the density matrices with rank $r \leq 4$;
and \textsl{ii.} vanishes after the finite time for the rank-8
density matrices. In future we like to investigate in more details
how the accuracy of the lower bound approximation for three-qubit
concurrence depends on the rank of the density matrix.

\begin{acknowledgments}
This work was supported by the DFG under the project No. FR 1251/13.
\end{acknowledgments}

\end{document}